\documentclass[12pt]{iopart}
\usepackage{graphicx}

\newcommand{\ket}[1]{{| {#1} \rangle}}

\begin{document}

\title{Susceptibility of Trapped-Ion Qubits to Low-Dose Radiation Sources}

\author{Jiafeng Cui$^1$, A. J. Rasmusson$^1$, Marissa D'Onofrio$^1$, Yuanheng Xie$^1$, Evangeline Wolanski$^1$, Philip Richerme$^{1,2}$}
\address{$^1$Indiana University Department of Physics, Bloomington, Indiana 47405, USA}
\address{$^2$Indiana University Quantum Science and Engineering Center, Bloomington, Indiana 47405, USA}

\date{\today}

\begin{abstract}
We experimentally study the real-time susceptibility of trapped-ion quantum systems to small doses of ionizing radiation. We expose an ion-trap apparatus to a variety of $\alpha$, $\beta$, and $\gamma$ sources and measure the resulting changes in trapped-ion qubit lifetimes, coherence times, gate fidelities, and motional heating rates. We found no quantifiable degradation of ion trap performance in the presence of low-dose radiation sources for any of the measurements performed. This finding is encouraging for the long-term prospects of using ion-based quantum information systems in extreme environments, indicating that much larger doses may be required to induce errors in trapped-ion quantum processors.
\end{abstract}

\maketitle

\section{Introduction}
 Trapped atomic ions are a leading candidate for scalable quantum computation, with long coherence times \cite{wang2021single}, near-background-free quantum readout \cite{noek2013high}, and the highest-reported single- and two-qubit gate fidelities \cite{gaebler2016high,ballance2016high}. Such fidelities exceed the $\approx 99\%$ minimum threshold required for fault-tolerant quantum computing \cite{knill2005quantum,raussendorf2007fault}, and several experiments have demonstrated the preparation, error detection, and manipulation of logically-encoded qubits \cite{nigg2014quantum,linke2017fault,egan2020fault}. Yet, it is currently unknown whether these advantageous properties and high fidelities of trapped ion quantum systems persist in the presence of ionizing radiation, as may be found in space or other extreme environments. Understanding the degree to which radiation-induced errors arise in ion-trap processors will be crucial for mitigating potential failure mechanisms of trapped-ion quantum protocols.
 
 Recently, experiments with superconducting qubits have found that ionizing radiation from small-scale sources \cite{vepsalainen2020impact} and from cosmic rays \cite{mcewen2021resolving} can limit the qubit coherence times and destroy quantum information stored throughout the chip. In both cases, it is believed that ionizing radiation generates phonons in the chip substrate, breaking Cooper-paired electrons and producing large quasiparticle densities which lead to qubit decoherence \cite{martinis2009energy,kozorezov2000quasiparticle}. Since such radiation events may lead to widespread correlated errors between qubits, they may be difficult or impossible to correct using standard fault-tolerant methods \cite{fowler2014quantifying}.
 
 To date, no comparable studies of radiation effects have been performed using trapped-ion quantum processors. Although Cooper-pair breaking and quasiparticle generation are not applicable to ion-based qubits, ions may instead be susceptible to alternative radiation-induced effects. For instance, most ionizing radiation contains enough energy to increase the charge state of trapped ions and thereby destroy the qubit \cite{heugel2015resonant}. Even if the qubit survives, the presence of high-energy x-ray or $\gamma$ photons may induce Stark shifts \cite{drake1996atomic} or energy level fluctuations which reduce the qubit coherence time. Furthermore, high-energy radiation has the potential to ionize background gases or release adsorbed atoms and photoelectrons from the vacuum chamber walls \cite{turner2008atoms}, which may lead to increased collisions or motional heating of the ions.
 
In this work, we study the effects of low-dose radiation on trapped-ion qubits. We first expose an ion-trap apparatus to an array of laboratory-scale $\alpha$, $\beta$ and $\gamma$ radiation sources to observe whether the ion trapping lifetime is reduced. In the presence of those same sources, we next set limits on the changes in qubit coherence time and single-qubit rotation fidelity during exposure. Finally, we investigate whether low-dose radiation leads to increased motional heating rates of trapped ions. In all cases, our measurements of radiation effects are performed while the ion-trap is in operation, rather than irradiating the trap and testing afterwards.

In Section \ref{expt} below, we introduce the ion species and ion-trap apparatus used in this work. Section \ref{rad} describes the $\alpha$, $\beta$, and $\gamma$ sources integrated with the apparatus and lists their activity and estimated irradiance at the ion position. Lifetime, coherence time, gate fidelity, and motional heating rate results are presented in Section \ref{results}, followed by concluding remarks in Section \ref{conclusion}.


\section{Experimental Apparatus}
\label{expt}
\begin{figure*}[t!]
\label{fig:iontrap}
\begin{center}
\includegraphics[width=8cm]{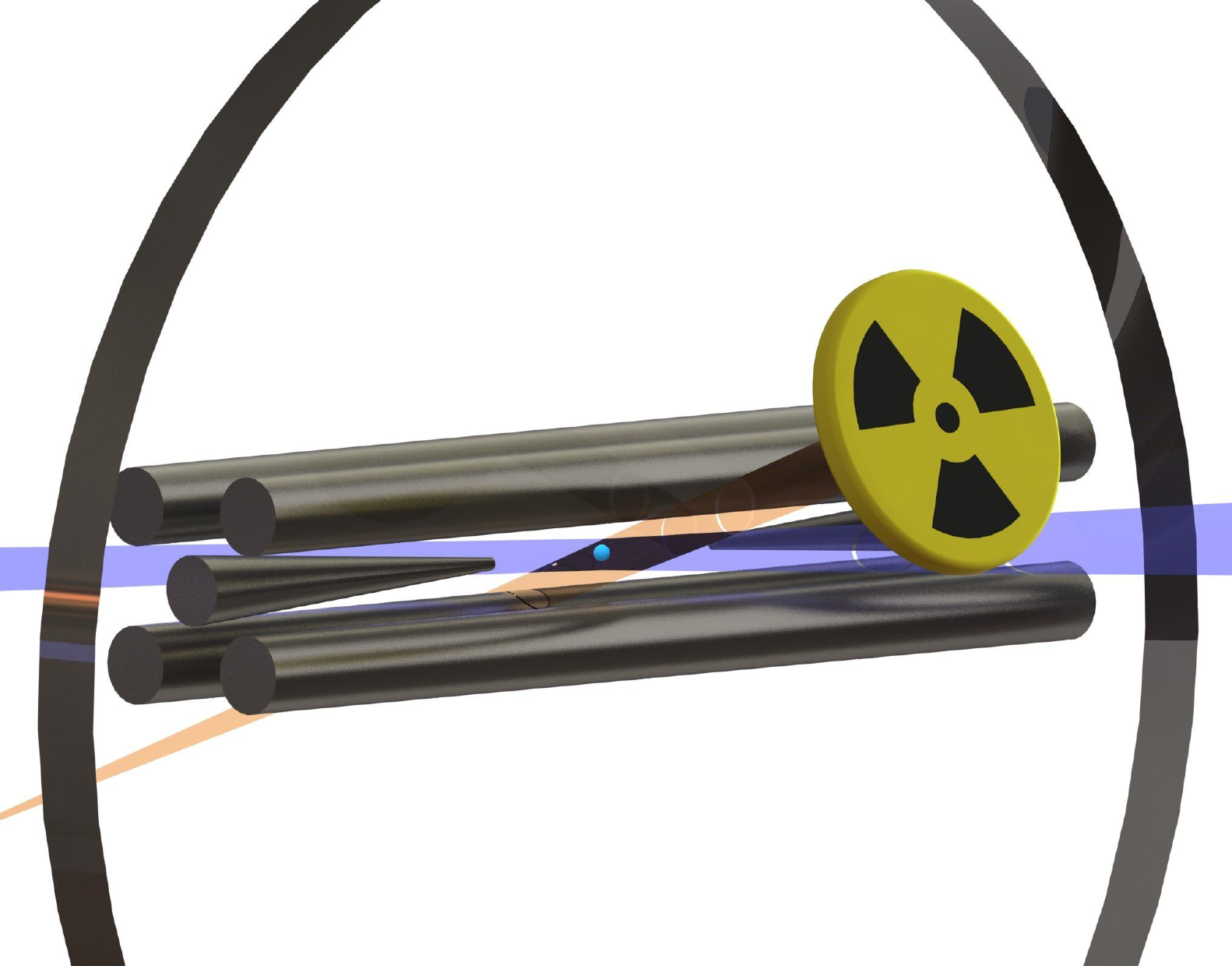}
\caption{Sketch of the experimental arrangement (not to scale). Ions are confined in a ``needle"-style rf trap housed inside a vacuum chamber. Laser beams (blue) are used for cooling and state-detection of the ions. Radiation (orange) emanates from a source outside the vacuum chamber and must pass through 4.65 mm of glass before interacting with the ions.}
\end{center}
\end{figure*}

The qubit used for these radiation testing experiments is encoded in the \mbox{$^2$S$_{1/2}~|F=0,m_F=0\rangle$} and \mbox{$|F=1,m_F=0\rangle$} hyperfine `clock' states of $^{171}$Yb$^+$ ions \cite{olmschenk2007manipulation}, denoted as $\ket{0}$ and $\ket{1}$, respectively. These states are separated by 12.6 GHz, addressable in the microwave regime, and are far enough separated in frequency to allow for high-fidelity state-dependent detection \cite{noek2013high}. In addition, these qubit levels are well-isolated from sources of decoherence, with effectively infinite $T_1$ relaxation times \cite{olmschenk2007manipulation} and demonstrated $T_2$ times in excess of one hour \cite{wang2021single}. For these reasons, $^{171}$Yb$^+$ has been a popular ion qubit of choice for groups performing quantum computation \cite{nam2020ground,pino2021demonstration,clark2021engineering} and simulation experiments \cite{monroe2021programmable}.

Ions are confined in a linear radiofrequency (rf) trap with ``needle"-style endcaps \cite{olmschenk2007manipulation,gulde2003experimental} (Fig. 1). Typical axial and radial trap frequencies for these experiments are $2\pi \times 390$ kHz and $2\pi \times 720$ kHz, respectively. The trap is housed within a vacuum chamber with pressures below \mbox{$10^{-10}$ Torr}, so that collisions with background gas particles are minimized \cite{wineland1998experimental}. Such low pressures are achieved by careful selection of the materials used inside the chamber, by stringent pumping and high-temperature bake-out processes, and by using ion pumps and non-evaporable getter pumps once in the ultra-high-vacuum regime. The walls of the vacuum chamber are made from 316L stainless steel, with three Corning 7056 glass viewports providing optical access to the ions. As shown in Fig. 1, radiation must also pass through one of the viewports before interacting with the ions.

For the $^{171}$Yb$^+$ qubits, the trapping, cooling, state initialization, and detection processes are all performed using lasers. Trapping begins by using lasers at \mbox{399 nm} and 369 nm to photoionize a beam of neutral Ytterbium atoms coming from a heated oven. Once the ion is created and trapped, lasers near 369 nm cool the ion to milliKelvin temperatures and initialize the qubit into the $\ket{0}$ state with $> 99\%$ fidelity \cite{d2020radial}. Detection of the trapped ion qubits is performed optically using standard state-detection fluorescence techniques \cite{noek2013high}, with the collected light imaged onto a photo-multiplier tube (PMT) or CCD camera. Finally, qubit state manipulation in these experiments is performed by broadcasting 12.6 GHz radiation, resonant with the qubit state separation, using a microwave horn just outside the vacuum chamber.

\section{Radiation Sources}
\label{rad}
We subject the ion trap hardware to an array of $\alpha$, $\beta$, and $\gamma$ radiation sources. Table 1 lists each of the isotopes used in these experiments, along with their radiation type, activity, and primary decay energies. Each source is encapsulated in a 1-inch ``button'' package, of U.S. Nuclear Regulatory Commission exempt quantity, and is mounted just outside the ion trap vacuum chamber (Fig. 1). To reach the ions, the radiation must pass through 45.5 mm of air, 4.65 mm of glass, and 10.5 mm of vacuum.

The presence of the glass vacuum window is expected to cause significant variations in the radiation dose at the ion, depending on the radiation type. For instance, it is known that $\alpha$-radiation can be stopped by a piece of paper, while thick lead shielding is often required to attenuate $\gamma$-rays \cite{turner2008atoms}. Consequently, the source activity alone is not sufficient to determine the radiation dose at the ion; interactions between the radiation and the vacuum window must be considered as well.

We estimate the irradiance at the ion for each source in Table 1, which is equivalent to the radiative energy flux passing through the trapping region inside the vacuum chamber. Several different numerical methods were implemented to estimate the attenuation of $\alpha$, $\beta$, and $\gamma$ particles through the vacuum window. Since $\alpha$ particles are positively charged, we simulated their trajectories using the Monte-Carlo based Stopping and Range of Ions in Matter (SRIM) code \cite{ziegler2010srim}. For the decay energies of $^{210}$Po and $^{241}$Am used in these experiments, a typical $\alpha$-particle is estimated to penetrate only $\sim20-30~\mu$m into the \mbox{4.65 mm} glass window, with negligible probability to pass through the full thickness. Similarly, $\beta$-attenuation was estimated using a Monte-Carlo simulation of electrons in solids (CASINO) \cite{hovington1997casino}, with non-negligible transmission probabilities found for only the highest-energy $\beta$-particles. In contrast, high transmission probabilities were found for $\gamma$-rays of the energies used in these experiments, as estimated using the NIST XCOM photon cross-section database \cite{berger2010xcom}.

\begin{table*}[t!]
\caption{Low-dose $\alpha$, $\beta$, and $\gamma$ sources integrated with the ion trap experimental apparatus. For each source, the activity and primary decay energies are listed, as well as the estimated irradiance at the location of the trapped ion. The presence of the vacuum window between the sources and the trapped ion shields essentially all $\alpha$ particles and most $\beta$ particles.}
\centering
\begin{tabular}{@{}lllll}
\br
 \textbf{Source} & \textbf{Type}  & \textbf{Activity } & \textbf{Energy (keV)} & \textbf{Irradiance (W/m$^{-2}$)} \\ 
\mr
Polonium-210 & $\alpha$ & 0.1 $\mu$Ci & 5300 & $\approx$ 0 \\
Americium-241 & $\alpha$ & 1 $\mu$Ci & 5490 & $\approx$ 0 \\
Strontium-90 & $\beta$ & 0.1 $\mu$Ci & 546 & $2\times10^{-15}$\\
Thallium-204 & $\beta$ & 10 $\mu$Ci & 764  & $2\times 10^{-8}$ \\
Cobalt-60 & $\beta$, $\gamma$ & 1 $\mu$Ci & $\beta$: 318; $\gamma$: 1170, 1330  & $\beta$: $1 \times 10^{-18}$; $\gamma$: $2\times 10^{-4}$\\
Cesium-137 & $\beta$, $\gamma$  & 5 $\mu$Ci & $\beta$: 512, 1170; $\gamma$: 662  & $\beta$: $4 \times 10^{-7}$; $\gamma$: $2\times 10^{-4}$\\
Cadmium-109 & $\gamma$ & 10 $\mu$Ci & 88 & $6 \times 10^{-5}$\\
Barium-133 & $\gamma$ & 10 $\mu$Ci & 81, 276, 304, 356, 384 & $3\times 10^{-4}$\\
\br
\end{tabular}
\end{table*}

\section{Results}
\label{results}
\subsection{Lifetime Measurements}
As an initial investigation of the effects of radiation on trapped ion qubits, we measure the trapping lifetime of ions exposed to radiation sources. The ``ion lifetime'' refers to the $1/e$ time for which an ion qubit remains confined within the rf trap in the absence of cooling mechanisms. If radiation induces fast depopulation of the ion trap, it may prove uncorrectable by both standard quantum error-correcting codes \cite{knill2005quantum,raussendorf2007fault,nigg2014quantum,linke2017fault,egan2020fault} or by more specialized codes which account for qubit loss \cite{stace2009thresholds,stricker2020experimental}.

Although ions have been confined in many systems for months, this requires continuous laser cooling which is forbidden while a quantum computation is in process \cite{monroe2013scaling}. Without active cooling, there is the potential for collisions to cause unmitigated ion heating and eventual loss of the qubit. Radiation effects may further increase the local background gas pressure and collision rate, as well as further ionize Yb$^+$, which in both cases would lead to reduced trapping lifetimes.

Our measurements find that ion lifetimes remain in excess of one second when exposed to every source of radiation listed in Table 1. For each experiment, one ion is initially Doppler cooled to 0.5 mK, confined without any cooling for 1 second, then illuminated to confirm its survival in the trap. Each experiment is repeated for 10 trials per data point. The $100\%$ measured survival probability at one second indicates that the true $1/e$ lifetime is longer than one minute in all cases. We note that for ion-trap experiments, one second is already orders of magnitude longer than the typical $\sim 1-10$ millisecond timescales of quantum computation and simulation studies \cite{linke2017experimental,richerme2014non}.


\subsection{Coherence time and Single-Qubit Gate Fidelity Measurements}
\begin{figure}[t!]
\centering
\includegraphics[width=12cm]{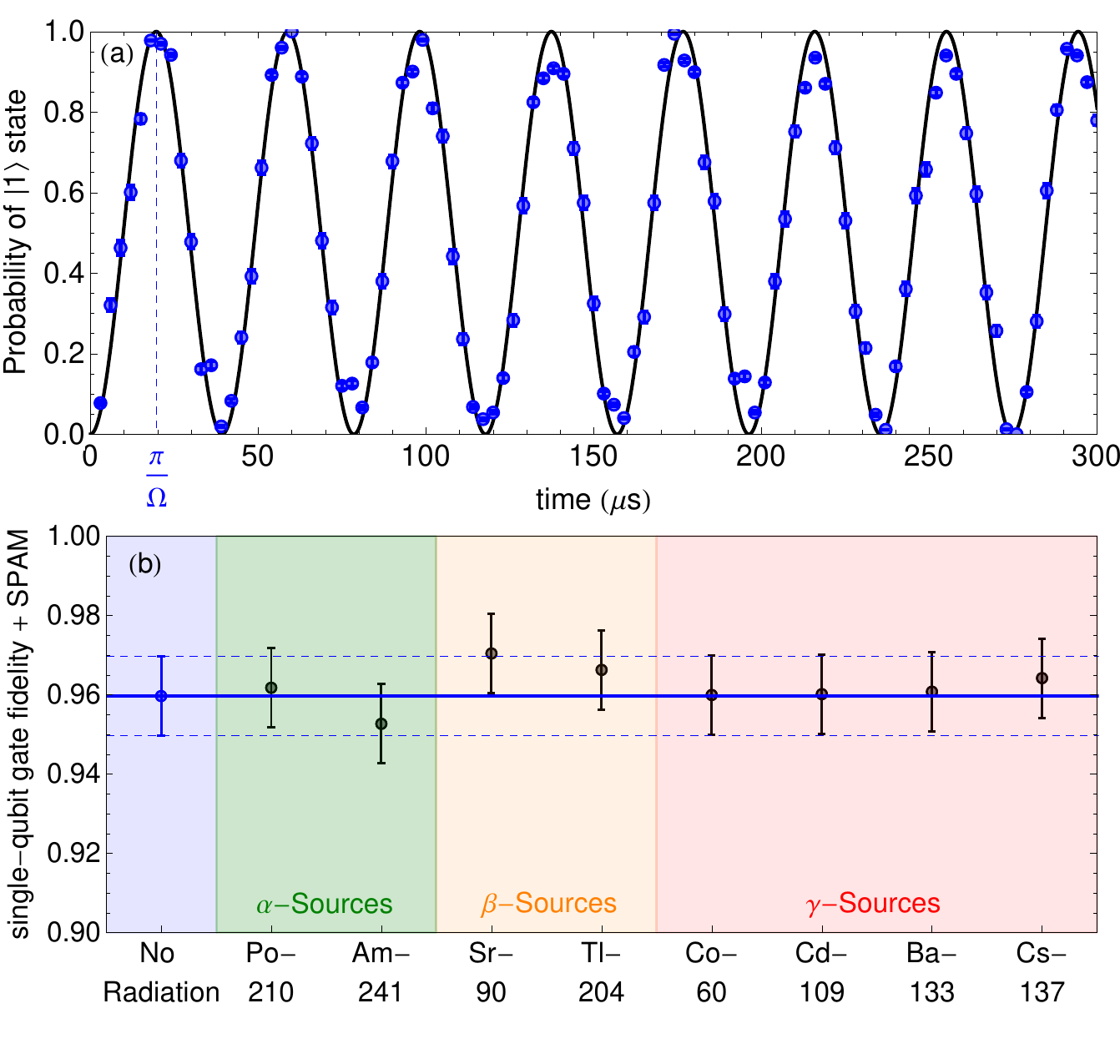}
\caption{(a) Oscillations between qubit states $\ket{0}$ and $\ket{1}$ when driven with microwave pulses at 12.6 GHz, with no radiation present. (b) The single-qubit $X$-gate fidelities (including all state-preparation and measurement errors) remain unchanged to within experimental error when various types of low-dose radiation are introduced. The blue solid line indicates the results of the control (no radiation) trial; blue dashed lines indicate 1 s.d. statistical measurement uncertainty.}
\label{fig:rabi}
\end{figure}

In this next round of experiments, we investigate whether (a) the coherence time of our system is sufficiently long to apply a single-qubit rotation in the presence of radiation, and if so, (b) whether the single-qubit gate fidelity is measurably impacted by the radioactive sources. Radiation-induced Stark shifts have the potential to cause uncontrolled variations in the phase evolution of the quantum bit \cite{drake1996atomic}, resulting in lowered $T_2$ coherence times and gate fidelities. If no decrease in fidelity is observed, then we can simultaneously conclude that (a) the coherence time is sufficiently long for single-qubit gate operations, and (b) radiation does not affect single-qubit rotations to within experimental error.

The single-qubit gates implemented here take the form of rotations around the $\hat{x}$ axis of the Bloch Sphere, $U(t)=e^{-i \sigma_x \Omega t/2}$, where $\sigma_x$ is the Pauli $X$ matrix and $\Omega$ is the Rabi frequency. Rotations are driven using microwaves resonant with the \mbox{12.6 GHz} frequency splitting between qubit levels. When the qubit is initialized in $\ket{0}$ and microwaves are left on continuously, the qubit state oscillates between the $\ket{0}$ and $\ket{1}$ states at Rabi frequency $\Omega\approx 25$ kHz, as shown in Fig. \ref{fig:rabi}(a). To estimate the average single-qubit $X$-gate fidelity, we apply the 12.6 GHz radiation for a time $t=\pi/\Omega$ and measure the population fraction transferred from the $\ket{0}\rightarrow\ket{1}$ state.

As described, such measurements underestimate the true single-qubit gate fidelity since they include the effects of State Preparation and Measurement (SPAM) errors. These errors account include the possibilities that the initial preparation is not purely $\ket{0}$, but contains some small fraction of $\ket{1}$, and that the measurement fidelity of distinguishing $\ket{0}$ from $\ket{1}$ is not 100$\%$. An independent characterization of our total SPAM error yielded $3.3 \pm 1.3\%$, which is the dominant source of infidelity for the measurements in Fig. \ref{fig:rabi}(b).

The measured single-qubit $X$-gate fidelities, including SPAM errors, is shown in Fig. \ref{fig:rabi}(b). Each experimental trial was repeated 10,000 times to keep quantum projection noise errors at the level of $\sim 10^{-2}$. To within experimental error, we observed no radiation-induced change during the combined operations of state preparation, single qubit rotations, and measurement when compared with our control trial. Independently, by performing multiple concatenated $X-$gates, we bound the maximum radiation-induced single-qubit fidelity error (in the absence of SPAM) at the $<0.3$\% level. We therefore conclude that (a) the coherence time of the trapped ion remained sufficient for single-qubit gate operations, and (b) if radiation effects were indeed present, they would be well under the $\sim$1\% error threshold needed for correction under fault-tolerant schemes.

\subsection{Heating Rate Measurements}
In a final set of experiments, we quantify the effects of radiation on the trapped ion temperature. Quantum entangling operations rely on cooling ions to near their ground state of motion \cite{molmer1999multiparticle}, such that their motion is quantized in a global harmonic oscillator potential. Since dissipative cooling is forbidden during quantum gate operations, ion heating during the computation may compromise the overall fidelity \cite{brownnutt2015ion}. If this heating rate is exacerbated by the presence of radiation, either through direct or induced collisions with background particles or by increased charge fluctuations on nearby surfaces, two-qubit gate fidelities will be negatively affected.

Our measurements of the ion temperature begin by Doppler-cooling the ion to $\approx0.5$ mK. The cooling laser is then turned off, and the ion is allowed to heat for 100 ms (much longer than the typical gate time of an ion-trap quantum computer \cite{linke2017experimental}). Finally, the temperature is re-measured after 100 ms such that the heating rate may be determined. We then repeat this sequence in the presence of all radioactive sources listed in Table 1.

Temperature measurements of a trapped ion are performed by observing its fluorescence as a function of detuning $\delta$ from resonance. Such resonance lineshapes have two primary contributions. The first is the power-broadened Lorentzian linewidth of the atom, given by $\Gamma'=\Gamma\sqrt{1+s}$, where $\Gamma=(2\pi)\times 19.6$ MHz is the natural linewidth of the Yb$^+$ 369 nm transition, and $s=0.3$ is the laser saturation parameter used in these experiments. The second major contribution to the linewidth comes from the Doppler-broadened temperature of the ions, which is a Gaussian lineshape with standard deviation $\sigma=\sqrt{k_B T/m\lambda^2}$, where $\lambda=369$ nm, $k_B$ is Boltzmann's constant, and $T$ is the ion temperature.

\begin{figure}[t!]
\centering
\includegraphics[width=12cm]{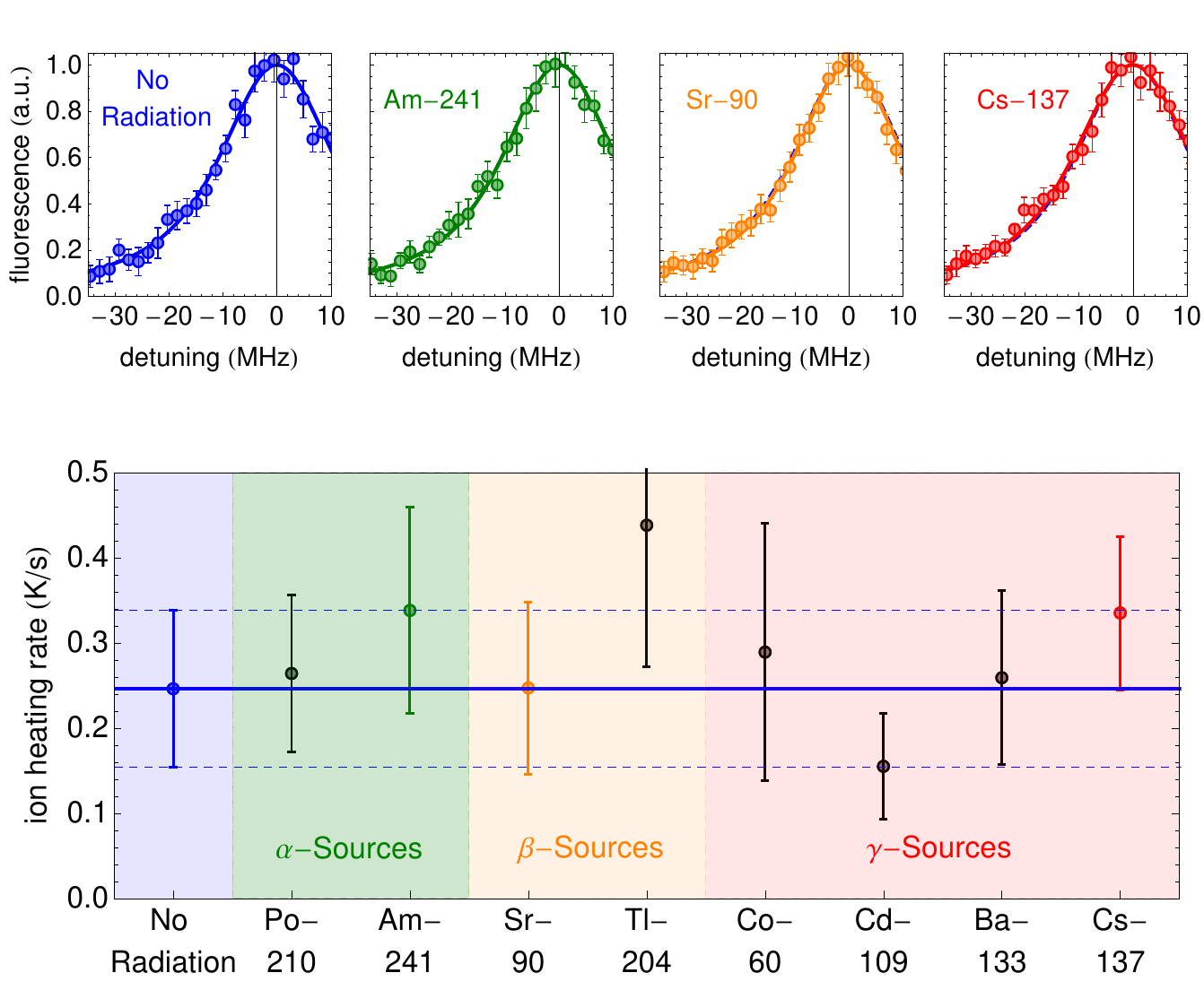}
\caption{(Insets): measured ion fluorescence vs. detuning from resonance. The width of the lineshape determines the ion temperature. (Bottom) The extracted ion heating rates for various radiation sources. No statistically significant deviation in the heating rate is observed compared to the no-radiation case. The blue solid line indicates the results of the control (no radiation) trial; blue dashed lines indicate 1 s.d. measurement uncertainty.}
\label{fig:heating}
\end{figure}

Given these two contributions to the linewidth, the fluorescence profile is most appropriately fit to a Voigt function, which is the convolution of a Gaussian lineshape $G$ and Lorentzian lineshape $L$:
\begin{equation}
V(\delta;\sigma,\Gamma')=\int_{-\infty}^{\infty}G(\delta';\sigma)L(\delta-\delta';\Gamma') d\delta'
\end{equation}
By fitting this lineshape to the measured fluorescence as a function of laser detuning, the only free parameter is the Gaussian width $\sigma$ which uniquely determines the ion temperature $T$.

Characteristic lineshapes for the control case (no radiation) and for $\alpha$, $\beta$, and $\gamma$ sources are shown in the top insets of Fig. \ref{fig:heating}. For the no-radiation case, the linewidth yields a fitted temperature of 25 mK after 100 ms of heating, corresponding to a heating rate of 0.25 K/s (6.0 quanta/ms). This lineshape is replicated as a blue dashed line in the other 3 inset panels but is almost entirely covered by the radiation-present data. For each isotope and dose of radiation we have fit a lineshape profile to extract a temperature and heating rate, plotting the summary of results in the bottom panel of Fig. \ref{fig:heating}.

We observe no statistically significant increase in the heating rates when the ion trap is irradiated with low-dose $\alpha$, $\beta$, or $\gamma$-sources. Likewise, we observe no increase in the background gas pressure at the $10^{-10}$ Torr level (which would increase the collisional heating rate), nor do we find that the ion shifts position in the trap due to unwanted charge accumulation (which would increase the electric field noise heating rate). We therefore conclude that much higher doses of radiation may be necessary to cause measurable increases in ion heating rates and their associated entangling gate infidelities.

\section{Conclusion}
\label{conclusion}
This work measured the \emph{in-situ} changes in ion-trap qubit lifetimes, coherence times, single-site rotation fidelities, and motional heating rates when exposed to an array of laboratory-scale $\alpha$, $\beta$ and $\gamma$ radiation sources. Since the ion trap was in operation during these measurements, the effects of radiation were tested on the ion qubit and surrounding trap hardware simultaneously. If radiation-induced errors were found to exceed fault-tolerant thresholds, it would indicate serious future challenges for the ability to perform large-scale quantum computations using unshielded ion-trap-based hardware. 

For the small-scale doses used in this study, we found no quantifiable degradation of ion-based qubits in the presence of radiation, for any of the measurements performed. This finding is an early first step for demonstrating the long-term prospects of using ion-based quantum information systems in space or other extreme environments. However, exposure to higher-dose sources will be required to fully quantify possible points of failure and guide future design requirements for system shielding. In addition, future work should also more accurately quantify the single-qubit gate errors using randomized benchmarking, and determine whether high-dose radiation induces correlated ion-qubit errors which cannot be easily corrected using standard fault-tolerant protocols. 

\ack
This work was supported by the U.S. Office of Naval Research, NSWC Crane, under Award $\#$N00164-20-1-1003. The IU Quantum Science and Engineering Center is supported by the Office of the IU Bloomington Vice Provost for Research through its Emerging Areas of Research program. The authors are grateful to Stephen Howell and Jonathan Dilger for stimulating discussions.

\section*{References}
\bibliographystyle{prsty}
\bibliography{main}
\end{document}